\documentclass[a4paper]{article}
\usepackage{graphicx} 
\usepackage[english]{babel}
\usepackage{lmodern}
\usepackage{microtype}
\usepackage{amsmath, amssymb, amsthm, amsfonts}
\usepackage{etoolbox, scrlayer-scrpage}
\usepackage{pdfpages}
\usepackage{multirow}
\usepackage{booktabs}
\usepackage{float}
\usepackage[hidelinks]{hyperref}
\usepackage{bm}
\usepackage{url}
\usepackage{subcaption}
\usepackage[a4paper]{geometry}
\usepackage[noblocks]{authblk}
\usepackage{rotating}
\usepackage[authoryear]{natbib}

\makeatletter
\patchcmd{\@maketitle}{\huge}{\Large}{}{}
\patchcmd{\abstract}{\quotation}{}{}{}

\newcommand{\shorttitle}{\@title}
\makeatother

\title{\textbf{Enhancing the use of family planning service statistics using a Bayesian modelling approach to inform estimates of modern contraceptive use in low- and middle-income countries}}
\author[1]{Shauna Mooney\thanks{shauna.mooney.2015@mumail.ie}}
\affil[1]{\textit{Hamilton Institute and Department of Mathematics and Statistics, Maynooth University, Ireland}}
\author[2]{Leontine Alkema}
\author[3]{Emily Sonneveldt}
\author[3]{Kristin Bietsch}
\author[3]{Jessica Williamson}
\author[1]{Niamh Cahill}
\affil[2]{\textit{Department of Biostatistics and Epidemiology,
University of Massachusetts Amherst, USA}}
\affil[3]{\textit{Avenir Health, Glastonbury, CT, USA}}

\begin{document}

\maketitle

\begin{abstract}

\textbf{Background:} Monitoring family planning indicators, such as modern contraceptive prevalence rate (mCPR), is essential for family planning programming.  The Family Planning Estimation Tool (FPET) uses survey data to inform estimates and forecasts of family planning indicators, including mCPR,  over time. However, reliance solely on surveys can lead to data gaps given that large-scale, population-health surveys are carried out on average every 3-5 years. 
Service statistics are a readily available data source that are routinely collected in conjunction with family planning service delivery. Various service statistics data types can be used to derive a family planning indicator called Estimated Modern Use (EMU). In a number of countries, annual rates of change in EMU have been found to be predictive of true rates of change in mCPR. However, it has been challenging to capture the varying levels of uncertainty associated with the EMU indicator across different countries and service statistics data types and to subsequently quantify this uncertainty when using EMU in FPET.

\textbf{Methods: }In this paper, we present a new approach to using EMUs in FPET, to inform mCPR estimates. The new approach is based on using annual EMU rates of change as input, and accounts for uncertainty associated with the EMU derivation process. The approach also accounts for additional country-type-specific uncertainty. We assess the EMU type-specific uncertainty at the country level, via a Bayesian hierarchical modelling approach.

\textbf{Results: }We present model validation results and anonymised country-level case studies to highlight the impact of including EMU data with uncertainty in FPET when estimating mCPR. The validation results illustrate improved predictive performance with the inclusion of EMUs compared to using survey data only. Case studies provide additional insights into how including EMU data affects mCPR estimates in different country contexts. Together, the validation results and case studies demonstrate that EMUs can help countries more effectively monitor progress toward their family planning goals.

\end{abstract}

\section{Introduction}

Family planning supports the fundamental right of individuals to choose the number and timing of their children. Access to family planning greatly enhances health outcomes for women and children and helps reduce poverty (\cite{Cleland2006}). To effectively track progress toward country-level family planning goals, it is essential for countries to monitor current trends in family planning indicators, such as contraceptive prevalence and unmet need for contraception, and forecast future trends (\cite{Stover2017}). This empowers countries to make data-driven, informed decisions towards achieving family planning goals.

The Family Planning Estimation Tool (FPET) is a Bayesian statistical model that is used to produce country-level estimates and forecasts of family planning indicators (\cite{Alkema2013,Cahill2018, kantorov2020, fpet_evolution}). One of the key family planning indicators used to monitor progress is the modern contraceptive prevalence rate (mCPR), defined as the proportion of women reporting that themselves or their partner currently uses at least one modern contraceptive method. 

While FPET traditionally relies on survey-based observations of family planning indicators, the intermittent nature of large-scale population-health surveys, conducted every 3-5 years on average, introduces data gaps. This intermittency poses a challenge to achieving data-driven model forecasts and estimates. To address this, we draw on routine health facility data, specifically family planning service statistics, which serve as a supplementary data source generated as a by-product of family planning service delivery.

Family planning service statistics are used to derive Estimated Modern Use (EMU), a family planning indicator that provides insight into modern contraceptive use (\cite{emutrack20, ss2emu_tool}). Despite potential biases in EMU data, studies have demonstrated their utility in FPET to inform mCPR estimates in the absence of recent survey data (\cite{Magnani2018, Cahill2021}). To account for such biases, rates of change in EMU data, which are assumed to be unbiased with respect to rates of change in mCPR, can be used to inform mCPR estimates and forecasts where survey data are absent.

However, while annual rates of change in EMU estimates have been found to be predictive of true rates of change in mCPR, it has been challenging to capture and quantify the varying sources of uncertainty at the country level associated with this indicator. We present a new approach to using EMUs to inform mCPR estimates in FPET, accounting for both uncertainty associated with the EMU derivation process and the unexplained errors in country-specific EMU data series. Effectively quantifying this uncertainty ultimately improves the use of EMU data in FPET and results in improved accuracy and reliability of the EMU-informed estimates, enabling better tracking of mCPR trends. 

There is growing interest in the use of non-standard data sources, such as service statistics, to provide frequent and up-to-date insights into population-level health indicators (\cite{Hung2020, routine_data_strategy}). This is especially useful when informing decision making, particularly where 'gold standard' data such as household surveys are collected intermittently, not keeping up with the timelines of initiatives, therefore increasing the risk of relying on outdated data. To address this issue, researchers have explored the value of using various forms of non-standard data. For instance, researchers established a spatio-temporal model to combine survey data with routine health data to estimate malaria risk in Rwanda (\cite{Semakula2023}). Research into the coverage of maternal and child health services using routine health data further illustrates the use of readily available data to inform indicators (\cite{Maiga2021, Agiraembabazi2021}). Additionally, routine health data can offer insights into healthcare performance during crises (\cite{Turcotte-Tremblay2023}). These methods can allow for more responsive decision-making, reducing the risk of relying on outdated information. Beyond the use of routine health data, mobile phone data has been used to predict the spatial spread of cholera (\cite{Bengtsson2015}), and social media data has been used to track migration patterns (\cite{Alexander2022}). The work presented in this paper, improving the use of family planning service statistics to refine estimates of a key family planning indicator such as mCPR, further advances progress in this field.

The remainder of this paper is structured as follows: we begin with background, including an overview of FPET. This is followed by a section on service statistics. Next, we present our exploratory analysis and methodology. We then outline the model validation results, discuss the impact of EMU inclusion, and showcase country-level case studies to demonstrate the impact of EMU inclusion across various scenarios. We conclude with a final discussion.

\section{Background}
\subsection{Estimating mCPR using the Family Planning Estimation Tool} 

FPET produces estimates and short-term forecasts of the modern contraceptive prevalence rate (mCPR) for women of reproductive age, by marital status (\cite{Alkema2013,Cahill2018, kantorov2020, fpet_evolution}). 
mCPR is defined as the proportion of women who are users of modern methods of contraception, including female and male sterilisation, male and female condoms, hormonal methods, vaginal barrier methods, standard days method, lactational amenorrhea method, and emergency contraception. Interest lies in mCPR among married women of reproductive age (MWRA), unmarried women of reproductive age (UWRA), and all women of reproductive age (AWRA).

\begin{figure}[H]
  \centering
  \includegraphics[width=1\textwidth]{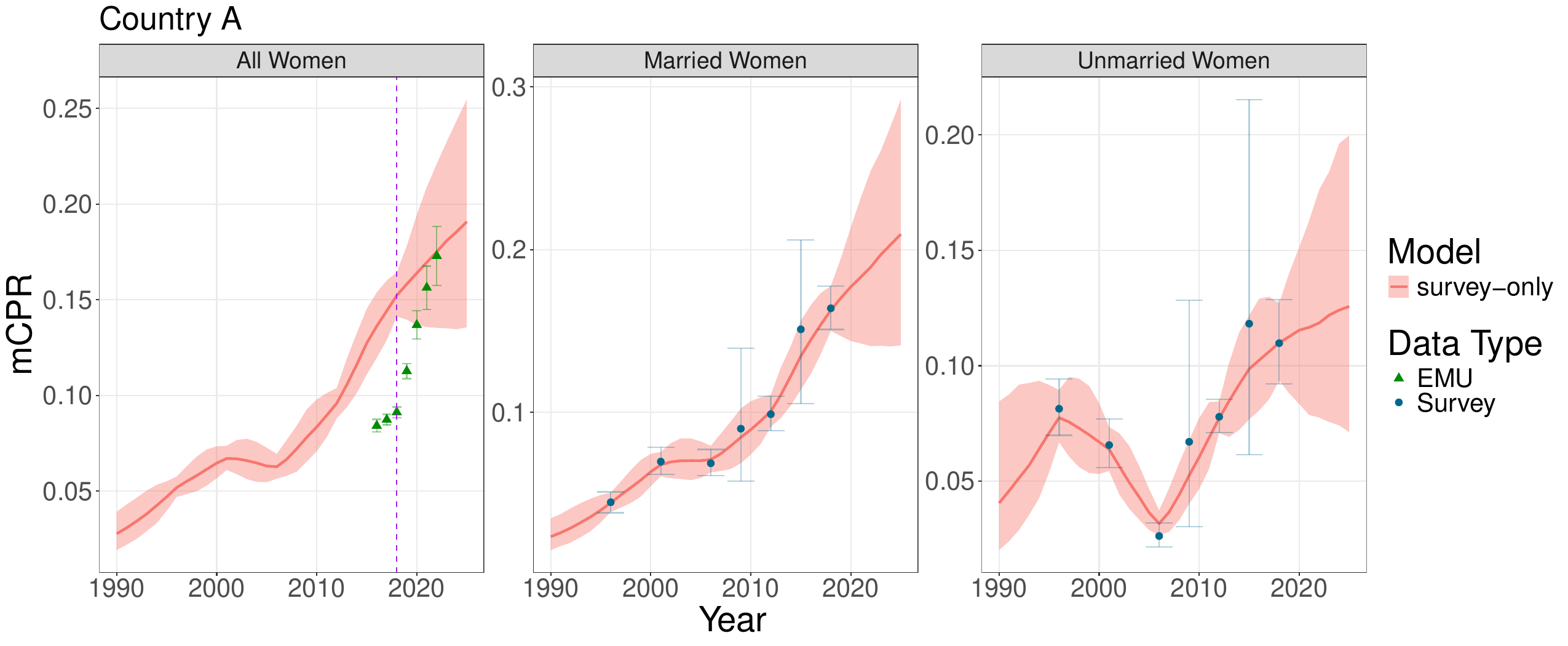}
  \caption{FPET estimates and forecasts of mCPR for AWRA, MWRA, and UWRA in Country A, using the survey-only model. The red solid line represents the median estimate of mCPR, with the shaded area indicating the 95\% credible interval. Green points show EMU data, which were not used to fit the model but are included for reference. Blue points represent the survey data used in the model fitting. The error bars represent the 95\% confidence interval associated with the data. The purple dashed line marks the year of the most recent survey.}
  \label{fig:countrya_so_mod_fit}
\end{figure}

FPET primarily produces estimates that are informed by survey data. The statistical model is based on a Bayesian hierarchical B-spline transition model to capture long- and short-term changes in family planning indicators over time, comparable to an ARIMA(1,1,0) model with level-dependent drift (\cite{fpet_evolution}). The survey data model, that captures how survey data is assumed to relate to the true family planning indicators, accounts for various types of errors associated with the data (\cite{alkema_temporal_2024}). Figure \ref{fig:countrya_so_mod_fit} illustrates FPET's survey-based estimates and forecasts of mCPR over time in a selected country (referred to as Country A) for AWRA, MWRA and UWRA. To ensure data confidentiality, we anonymise each country-level case study presented in this paper.

\subsection{Using service statistics data}

Service statistics provide a readily available, supplementary data source to survey data that can be used to inform estimates of mCPR in the absence of recent surveys. This is illustrated for Country A in Figure~\ref{fig:countrya_so_mod_fit}. In Country A, the latest survey occurred in 2018 (indicated by the dashed vertical line). Service statistics data extend to 2022 and could offer more up-to-date insights. EMUs, derived from Country A's service statistics data for AWRA, are added to the same figure and suggest a faster rate of change since 2018, as compared to the survey-based forecasts. The open question is how to use such data appropriately, including considerations of inherent biases and uncertainties.

\section{Service statistics}

\subsection{Health Management Information Systems}

Service statistics are collected via health management information systems (HMIS), which have been implemented in many low- and middle-income countries to routinely collect and manage data on healthcare services delivered at facilities. The most common being the DHIS2 platform, an open-source HMIS used by many FP2030 pledging countries, that is, countries that pledged to take specific actions to expand access to voluntary, rights-based contraception (\cite{measure_evaluation_2017, dhis2, fp2030}). We consider service statistics collected after the nationwide implementation of DHIS2.

\subsection{Service Statistics and Estimated Modern Use} \label{ss section}

There are four types of family planning service statistics:
\begin{enumerate}
    \item Number of contraceptive commodities, for example pill packets and intrauterine devices, distributed to clients (EMU-clients).
    \item Number of contraceptive commodities distributed to facilities (EMU-facilities).
    \item Number of times clients interacted with a provider for contraceptive services (FP visits).
    \item Number of current contraceptive users of any method including those who are still using longer acting methods that were received in previous years (FP users).
\end{enumerate}

The process of using service statistics to calculate EMUs was developed by Track20, a project dedicated to collaborating with and monitoring progress of countries involved in the FP2030 initiative (\cite{ss2emu_tool}). Further details of this calculation can be found in \cite{mooney_emu1}. EMUs are classified into one of the four data types, based on the service statistics used in their calculation. All data in this analysis were sourced from service statistics collected in 2023. The EMU dataset includes 344 observations from 23 countries. The volume and types of data vary by country; for instance, some countries have data from multiple EMU data types, while others have data from only one.

\section{Exploratory Analysis}

\begin{figure}[H]
  \centering
  \includegraphics[width=1\textwidth]{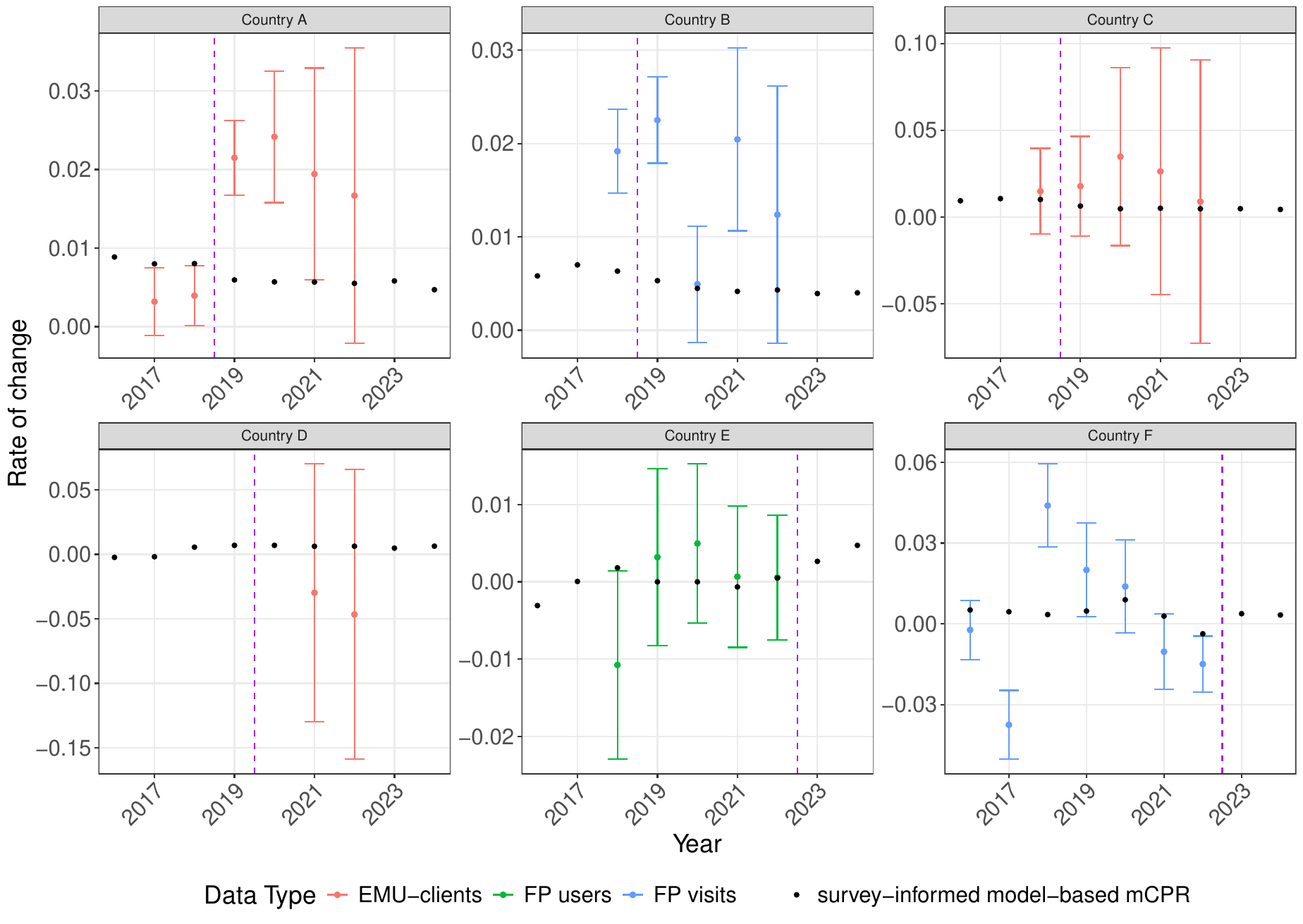}
  \caption{Rates of change in EMU and survey-informed model estimates of mCPR over time for Country A, Country B, Country C, Country D, Country E and Country F. Point colors represent EMU data type for each country. The error bars represent the 95\% confidence interval associated with the data. The purple dashed line marks the year of the most recent survey.}
  \label{fig:example_roc emu}
\end{figure}

Figure~\ref{fig:countrya_so_mod_fit} introduced survey-based model estimates of the level of mCPR in Country A, along with EMUs. When including EMUs in FPET, we consider EMU annual rates of change. Using survey-based model estimates of the level of mCPR, we can also derive survey-based annual rates of change in mCPR. This enables us to compare the rates of change observed in EMUs with those based on survey-only mCPR estimates. Figure \ref{fig:example_roc emu} presents examples of EMU rates of change and survey-based mCPR rates of change for six countries and three EMU data types. Error bars provide insight into the uncertainty associated with each EMU rate of change observation, derived during the calculation process (\cite{mooney_emu1}).

Observation-level uncertainty varies significantly across countries, as shown in Figure \ref{fig:example_roc emu}. For instance, EMUs in Country C and Country D tend to have higher uncertainty compared to the other countries shown. Uncertainty also fluctuates within each country, for example, in Country A, Country B, and Country C, the plot illustrates increasing EMU uncertainty over time. Generally, when taking into account observation-specific uncertainty, we can see that EMU can capture survey-informed mCPR rates of change, but there are notable deviations to be considered within countries and across types. 

Figure \ref{fig:example_roc emu} provides insight into how well EMU can capture survey-informed mCPR rates of change in each country. Given that this is better evaluated during survey-informed years, we focus on observations prior to the most recent survey year, represented using the purple dashed line. In Country A, survey-based rates of change in 2017 and 2018 are at the upper bounds of the 95\% confidence interval associated with EMU-based rates of change. Countries B and C both have only one observation prior to the survey year. While Country C's observation captures the survey-informed estimate, Country B's does not. In Country D, the absence of EMUs prior to the most recent survey makes it difficult to evaluate how well EMUs track mCPR trends. EMUs in Country E capture trends in survey-informed mCPR reasonably well. Initially in Country F, there is considerable variation in the EMU trends, however, more recent EMU observations show improvement in capturing survey-informed mCPR rates of change. 

\begin{figure}[H]
    \centering
    \includegraphics[width=.9\textwidth]{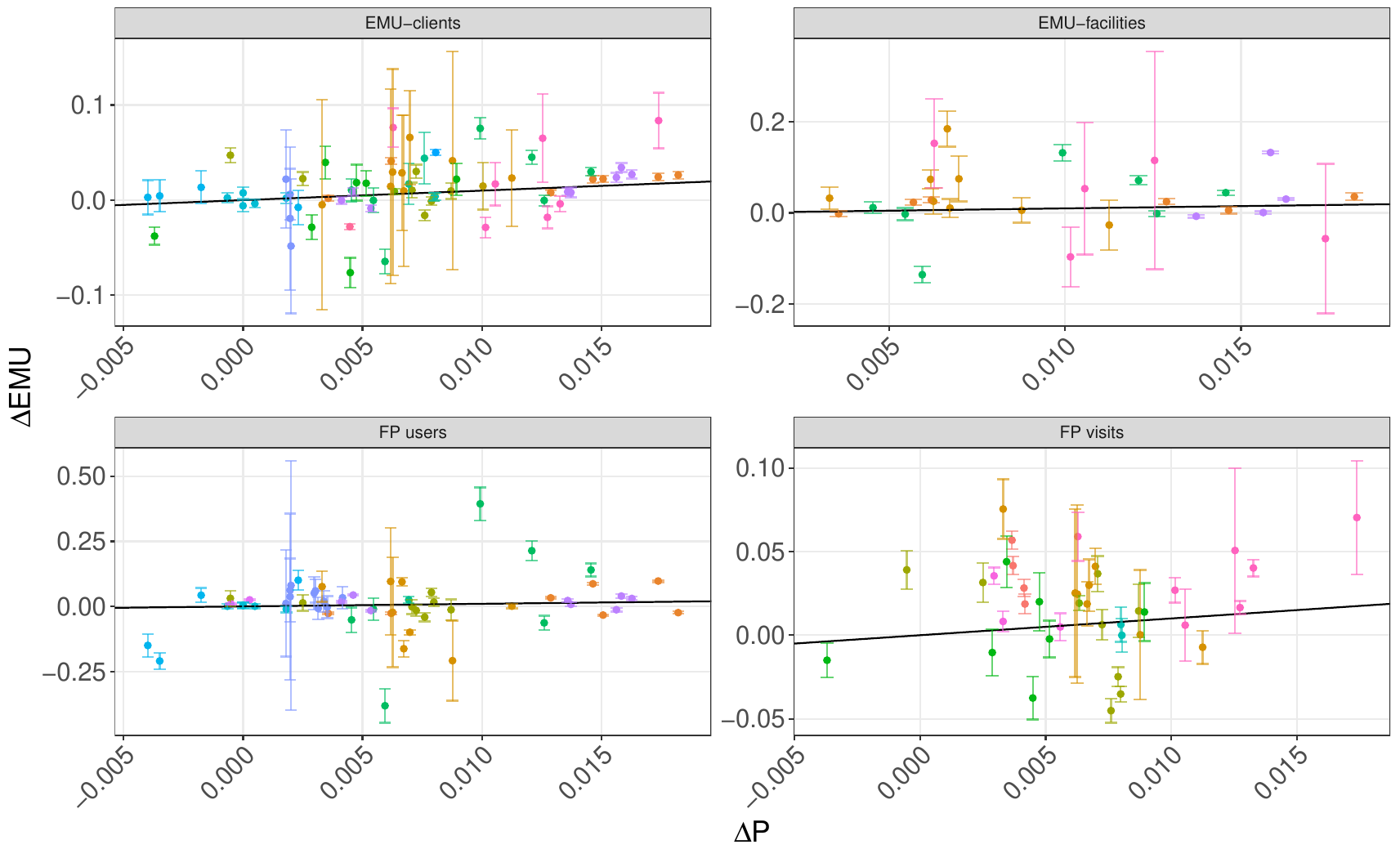} 
    \caption{Scatterplot of changes in EMU ($\Delta EMU$) versus changes in survey-informed model mCPR estimate ($\Delta P$) by EMU data type, coloured by country. The error bars represent the 95\% confidence interval associated with the data. Black lines represent the identity line ($\Delta EMU = \Delta P$).}
    \label{fig:delta_emu_vs_delta_p}
\end{figure}

Figure \ref{fig:delta_emu_vs_delta_p} provides an overview of rates of change in EMU ($\Delta EMU$) versus rates of change in survey-informed mCPR ($\Delta P$), for all data observed prior to the most recent survey in all countries in the database. The plot reveals that EMU-facilities and FP users display greater variation than EMU-clients and FP visits. In general, the changes in EMU data are dispersed around the identity line. However, differences across countries are notable. For example, in the FP users plot, the country represented in green shows some of the most extreme variations. Similarly, in the EMU-facilities plot, the country represented in pink displays more variability than the country represented by the orange data points. By ensuring our model incorporates cross-country variation in addition to type-specific variation, we can better capture these relationships. This further motivates our goal for an EMU data model in FPET that captures country-specific contexts, specifically, how well EMUs can predict trends in mCPR, while also accounting for observation-specific uncertainty.

\section{Methodology}

We introduce an EMU data model for use in FPET to inform estimates and forecasts of mCPR. This model incorporates recent advancements in EMU uncertainty quantification, using observation-specific uncertainties derived during the EMU calculation process. In addition, we capture additional country, type specific uncertainties informed by a set of hierarchically estimated variance hyperparameters. The Bayesian Hierarchical model used to estimate these hyperparameters is outlined in the next section. 

\subsection{EMU data model}

For the EMU data model within FPET, we define $\Delta{z}_{c,d,t}$ as the annual rate of change in EMU for country $c$, data type $d$, and year $t$, and $s_{c,d,t}$ as the corresponding standard deviation. Then for $\Delta{z}_{c,d,t}$ the model is specified as follows:

\begin{align}
    \Delta {z}_{c,d,t}|\Delta \rho_{c,t}, s_{c,d,t}, \sigma_{c,d} &\sim \mbox{N}(\Delta \rho_{c,t}, s_{c,d,t}^{2} + \sigma_{c,d}^2),
\end{align}

where $\Delta \rho_{c,t}$ represents the annual rate of change in mCPR for country $c$, at time $t$, observation-specific uncertainty is a data input, and $\sigma_{c,d}^2$ is an unknown variance parameter for country $c$ and data type $d$. This model assumes that trends in EMUs directly correspond to trends in true mCPR, with deviations away from this relationship attributed to observation-specific uncertainty and country- and type-specific uncertainty. 

The hierarchical model for the country-type variance is as follows:
\begin{align}
\log(\sigma_{c,d})|\theta_d, \tau \sim \mbox{N}(\theta_d, \tau^2),
\end{align}
where the log formulation for $\sigma_{c,d}$ ensures the necessary positivity constraint, $\theta_d$  refers to the type-specific mean and $\tau$ to the standard deviation of log-transformed standard deviation parameters. The hierarchical model results in estimates for $\sigma_{c,d}$ that are based on data from the specific country-type setting where available, with shrinkage towards type-specific means in data-limited settings.

\subsection{Estimating hyperparameters}

To estimate the hyperparameters $\theta_d$ and $\tau$ used in the EMU data model, we fit the Bayesian hierarchical model to training data from multiple countries and data types, that leverages information sharing across countries and data-types. Specifically, we use training data comprising of all EMU data and mCPR estimates available prior to the most recent survey, denoted as $\Delta {z}^{*}_{c,d,t}$ and $\Delta \rho^{*}_{c,t}$. This training dataset includes 203 observations of EMU and mCPR estimates across 19 countries.

The hierarchical model allows the estimates of  $\theta_d$  and $\tau$ to be informed by data from all countries in the training dataset. This means that even if there is limited data from a particular country, the estimates benefit from information being pooled across all countries in the dataset. The observed differences between rates of change in EMU and rates of change in mCPR in the training dataset, along with observation-specific uncertainties, directly inform our estimate of data-type specific log standard deviation ($\theta_d$) and cross-country variance ($\tau$) as follows:

\begin{align}
   \Delta {z}^{*}_{c,d,t} - \Delta \rho^{*}_{c,t}| \sigma_{c,d} &\sim \mbox{N}(0,  s_{c,d,t}^{2} + \sigma_{c,d}^2),
\end{align}
where, as before, 
\begin{align}
\log(\sigma_{c,d})|\theta_d, \tau \sim \mbox{N}(\theta_d, \tau^2).
\end{align}
Priors for $\theta_d$ and $\tau$ are specified as a Normal distribution and a half-Cauchy distribution, respectively (\cite{Gelman2006, Polson2012}). 
\begin{align}
    \theta_d \sim \mbox{N} (0,2^2),
\end{align}
\begin{align}
    \tau \sim \mbox{C}^{+}(0,1).
\end{align}

\subsection{Inclusion of EMU data in FPET}
FPET is used to produce estimates for a particular country, using data from that country alone (\cite{New2017, fpetrpackage}). The EMU data model used in FPET is the one presented above, using point estimates of the hyperparameters:
\begin{align}
   \log(\sigma_{c,d})|\hat{\theta}_d, \hat{\tau} \sim \mbox{N}(\hat{\theta}_d, \hat{\tau}^2),
\end{align}
where $\hat{\theta}_d$ is the estimated overall type-specific uncertainty and $\hat{\tau}^2$ is the estimated across country variance, estimated from the training data.

\section{Results}
\subsection{Estimates of the Bayesian hierarchical model hyperparameters}
Estimates of $\theta_d$ are summarised in Table \ref{tab:sigma_summary}. The smallest standard deviation (SD) estimate, $\hat{\theta}_d$, is associated with EMU-clients data, at -4.06 (95\% credible interval (CI): (0.01, 0.03) on the original scale), while the largest SD estimate is associated with the EMU-facilities data type, at -2.77, on the log scale  (95\% CI: (0.03, 0.10)). SD estimates for FP visits and FP users are -3.56 (95\% CI: (0.01, 0.06)) and -3.10 (95\% CI: (0.02, 0.08)), respectively. The estimate of $\hat{\tau}$, which captures cross-country variation, is 0.84. 

\begin{table}[H]
\centering
\caption{Summary of data-type specific standard deviation estimates, ($\hat{\theta}_d$), posterior standard deviations (SD($\hat{\theta}_d$)) on the log scale, and the 95\% credible intervals (CI) for $\hat{\theta}_d$ back-transformed to the original scale.}
\begin{tabular}{lccccc}
\toprule
Data type   & N   &  $\hat{\theta}_d$ & SD($\hat{\theta}_d$) & 95\% CI for $\exp(\hat{\theta}_d$) \\
\midrule
EMU-clients (d = 1)   & 73   & -4.06 & 0.27 & (0.01, 0.03) \\
EMU-facilities (d = 2)  & 30   & -2.77 & 0.41  & (0.03, 0.10)  \\
FP visits (d = 3)      & 60   & -3.56 & 0.35  & (0.01, 0.06) \\
FP users (d = 4)      & 40   & -3.10 & 0.31  & (0.02, 0.08) \\
\bottomrule
\end{tabular}
\label{tab:sigma_summary}
\end{table}

\subsection{Global findings: Validation results}

To evaluate model performance, we use an out-of-sample, leave-one-out validation exercise. In this context, this process involves excluding the most recent survey observation for each country during model fitting and using these excluded observations as test points. This validation method is intended to replicate a typical use case of the model. Performance was measured using several metrics including coverage and prediction errors. We evaluated both MWRA and UWRA model results for what we will term the survey-only model and the survey+EMU Model.

In cases where multiple types of service statistics are available, there is a data type deemed most appropriate for use in FPET during the data review process (\cite{mooney_emu1}). When evaluating the impact of the updated EMU model framework on performance and estimates, we focus specifically on results using the selected data type for each country.

Table \ref{tab:val_summary} presents model validation results of mCPR obtained using the survey-only model and the survey+EMU model, for MWRA and UWRA . This provides an overview of the validation results, highlighting coverage, mean error (ME), mean absolute error (MAE), and root mean square error (RMSE). The ME indicates the average bias in model predictions, with positive values reflecting under-prediction. The MAE measures the average magnitude of the errors in the model's predictions, indicating how far the predictions are from the test observation, regardless of direction. The RMSE gives insight into the variation of the error terms.

Across all metrics the survey+EMU model consistently outperforms the survey-only model. For MWRA, incorporating the EMU data model reduces the ME from 0.3 to 0.1, indicating a reduction in bias when including EMU. Since ME indicates the overall bias of predictions, that is,  whether the model systematically overestimates or underestimates, a ME closer to zero suggests that the model’s predictions are more balanced, with less systematic bias. Improvement in model performance is further supported by a reduction in MAE from 2.9 to 2.8, indicating errors of smaller magnitude. Additionally, the RMSE decreases from 3.7 to 3.5, indicating a reduction in the variability of the errors. Coverage remains at 95.7\% for both models.

Predictability for UWRA estimates also improve with the inclusion of the EMU data, as reflected by the reduction in ME from 0.2 to -0.01, which points to reduced bias. The RMSE for UWRA decreases slightly from 2.9 to 2.8, highlighting a modest improvement in prediction accuracy. Coverage increases from 90.9\% to 95.5\%, indicating an improvement in the models ability to accurately project the test survey observation within the uncertainty bounds. 

\begin{table}[H]
\centering
\caption{Summary of validation results for survey-only and survey+EMU models, for MWRA and UWRA mCPR estimates, highlighting coverage, mean error (ME), mean absolute (MAE) and root mean square error (RMSE).}
\begin{tabular}{llrrrrr}
\toprule
Marital Status & Model & N & Coverage & ME & MAE & RMSE  \\ 
\midrule
Married    & Survey-only & 23  & 95.7\% & 0.3 & 2.9 &  3.7  \\ 
Married    & Survey+EMU & 23 & 95.7\% & 0.1 & 2.8 & 3.5  \\ 
Unmarried  & Survey-only & 22 & 90.9\%  & 0.2 & 2.3 & 2.9 \\ 
Unmarried  & Survey+EMU & 22 & 95.5\% & -0.01 & 2.3 & 2.8 \\ 
\bottomrule
\end{tabular}
\label{tab:val_summary}
\end{table}

\subsection{Global findings: Impact of inclusion of EMU in FPET}

We evaluated the impact that incorporating the EMU data model into FPET has on mCPR estimates and forecasts. We compared mCPR estimates for 2023 obtained using the survey-only model to those derived from integrating both the survey and EMU data models. It is important to note that, particularly due to country-level variability, the overall uncertainty associated with EMUs can be substantial. As such, we expect that in some settings, the inclusion of EMUs may have minimal impact, which demonstrates one of the model's strengths. We illustrate country-level case studies in the next section that highlight this. 

\begin{figure}[H] 
    \centering
    \includegraphics[width=\textwidth]{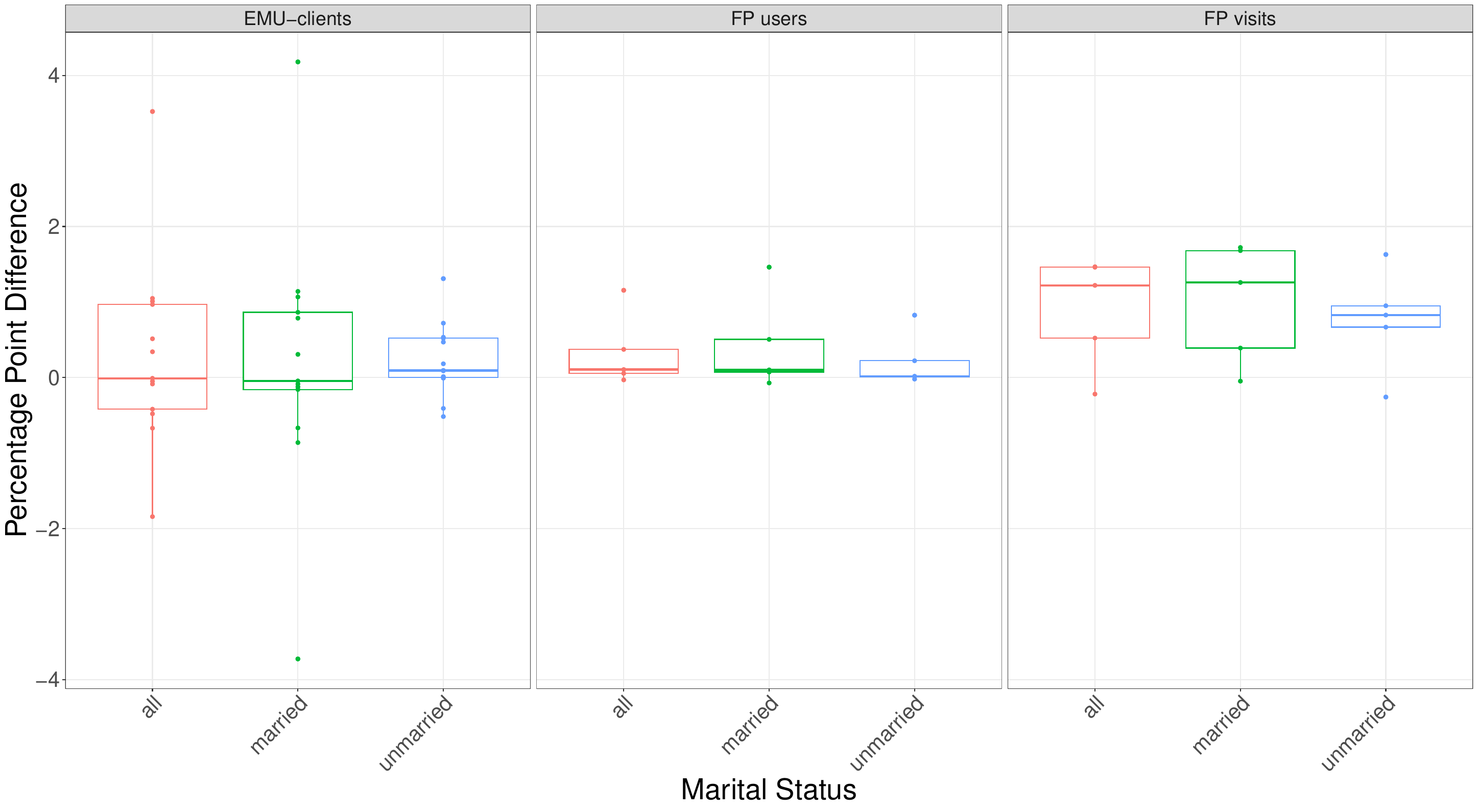}
    \caption{Boxplot of point estimate differences by marital status and data type, where the difference is defined as the mCPR estimate obtained with EMUs minus the mCPR estimate without EMUs (survey-only).}
    \label{fig:point_boxplot}
\end{figure}

Figure \ref{fig:point_boxplot} illustrates the differences in mCPR point estimates, by percentage points, across all countries in the database, categorized by marital status and EMU data type. A positive difference in mCPR indicates an increase when the EMU data is included compared to when only surveys are used. The use of EMU-clients EMUs results in the largest impact overall, observing a maximum increase in mCPR of 3.5 percentage points (pp), 4.2pp and 1.3pp, for AWRA, MWRA and UWRA  estimates respectively. FP users largest impact on point estimates of mCPR was 1.2pp, 1.5pp and 0.8pp on AWRA, MWRA and UWRA respectively. In terms of the use of FP visits, the largest change in point estimates was 1.5pp, 1.6pp, and 1.7pp; having the most impact on UWRA mCPR estimates in 2023. All data types show a positive median difference in mCPR estimates for UWRA, suggesting that supplementing the survey model with the EMU data model generally increases mCPR estimates for UWRA. EMU-clients data has the most variation in terms of point estimate differences, ranging from a decrease of 3.7pp to an increase of 4.2pp when considering MWRA estimates for example. The median point estimate difference when considering the use of FP visits is 1.2pp, 1.3pp and 0.8pp for AWRA, MWRA and UWRA, with the plot highlighting that these are the highest median point estimate differences across all data types. In terms of uncertainty with respect to model estimates, measured by the width of credible intervals, the inclusion of EMUs has no substantial effect.

\subsection{Country-level case studies}

\begin{figure}[H]
  \centering
  \begin{sideways}
    \includegraphics[width=1.5\textwidth]{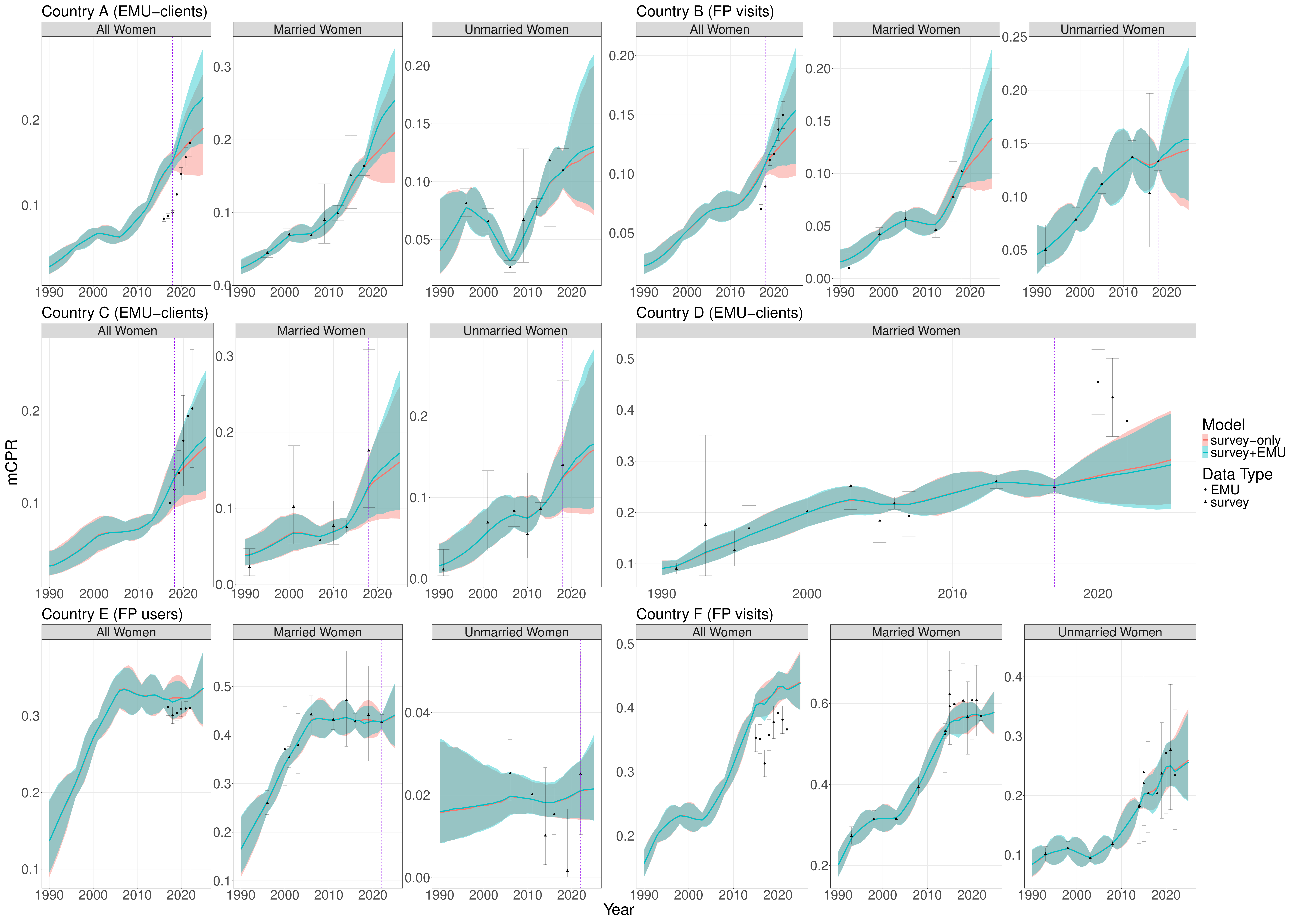}
  \end{sideways}
  \caption{Estimates of mCPR for MWRA, UWRA, and WRA in Country A, Country B, Country C, Country D, Country E, and Country F. Solid lines indicate median estimates of mCPR, shaded regions show the 95\% credible intervals. Results in red highlight the use of the survey-only model. Results in blue highlight the use of surveys and EMU. The error bars represent the 95\% confidence interval associated with the data. The purple dashed line marks the year of the most recent survey.}
  \label{fig:case_studies_mod_fits}
\end{figure}

To illustrate the impact of including the EMU data model at the country-level, we provide case studies from six countries in the EMU 2023 database. Figure \ref{fig:case_studies_mod_fits} highlights model results obtained using the survey-only and survey+EMU models for the six case study countries previously introduced in Figure \ref{fig:example_roc emu}. Results for all remaining countries can be found in the appendix.

When evaluating the impact that the inclusion of EMUs have on mCPR estimates at the country-level, there are three key components to consider. One key factor is observation-level uncertainty: when uncertainty is high, the impact on model results tends to be smaller. Another important consideration is how well EMUs align with or reflect survey-based mCPR estimates in a given country. Figures \ref{fig:example_roc emu} and \ref{fig:delta_emu_vs_delta_p} highlight the differences in both observation-level uncertainty, and how well EMUs align with mCPR across countries. Finally, the interval since the last survey is crucial: if a recent survey is available, the inclusion of EMUs will have little effect, while a longer interval can result in EMU inclusion having larger effect on estimates. 

Country A provides an example of a context where observation-level uncertainty associated with EMUs is low, as illustrated by narrow error bars, and EMUs show improvement in tracking mCPR during survey-informed years. As a result, the inclusion of EMUs have substantial impact on mCPR estimates when compared to the survey-only estimates. Recent EMU trends appear to be increasing more rapidly than the model-based mCPR estimates, potentially indicating a recent uptake in modern contraceptive use that has not been captured by a survey yet. When examining the 2023 model results, we observe increases of 3.5pp, 4.2pp and 0.7pp for AWRA, MWRA and UWRA compared to using the survey only model. 

Country B provide examples of contexts where observation-level uncertainty associated with EMUs is low, as illustrated by narrow error bars, however we don't have any EMU data before during survey-informed years to indicate EMUs are tracking mCPR well. In Country B, recent EMU trends appear to be increasing quicker than the model-based mCPR estimates, suggesting a recent increase in modern contraceptive use that has yet to be reflected in survey data. Point estimates of mCPR increase by 1.5pp, 1.7pp and 1pp for AWRA, MWRA and UWRA when including EMUs compared to the survey-only model. 

Country C presents a context with limited data prior to the most recent survey. However, the single available observation indicates that EMUs were tracking mCPR trends well during the limited, survey-informed period. Observation-level uncertainty associated with EMUs is increasing, as shown by widening the error bars, which reduces the impact of more recent EMU values. That said, including EMU results leads to a larger estimated increase in mCPR in the years since the most recent survey, compared to using surveys alone. Specifically, increases of 1pp for AWRA, 1.1pp  for MWRA, and 0.7pp for UWRA are observed compared to the survey-only model. 

In Country D, there is no EMU data available prior to the most recent survey. EMU data are collected for MWRA, and as such, these data are used to inform MWRA mCPR estimates in FPET. Unlike the previous case studies, EMU data show a decline in contraceptive use in the years since the most recent survey. This impacts model estimates in 2023 with a 0.8pp decrease in mCPR when including EMU data when compared to the survey-only model results.

We use Country E and Country F as illustrative examples to demonstrate the impact of using EMU data in contexts where there has been a recent survey (both countries conducted surveys in 2022). As expected, the inclusion of EMUs has minimal effect on mCPR estimates in these case studies.

\section{Discussion}

This paper introduces an EMU data model framework designed to directly inform mCPR estimates and forecasts within FPET in the absence of recent survey data. Our approach uses the latest advancements in the EMU calculation process, allowing us to incorporate a decomposition of EMU uncertainty into FPET for the first time. The decomposition includes uncertainties at the observational level as well as country, type-specific uncertainties. 

By taking into account observation-specific uncertainty in the EMU data model, observations with larger errors have less impact when used to inform mCPR estimates and forecasts at the country level. Previously, this type of uncertainty was not captured as EMUs were presented as point estimates. The results for six case study countries, highlight the uncertainty variation both across countries and over time within each country. This underscores the importance of incorporating observation-specific uncertainties when modelling EMU in FPET, as it allows for a more accurate representation of EMU rates of change.

In addition, the uncertainty decomposition provides flexibility in handling overall country-type specific uncertainty variations across countries. Previously, the EMU data model uncertainty was assessed solely by data type, which could lead to overdispersion in some countries and under-dispersion in others (\cite{Cahill2021}). The updated approach is more nuanced and can reduce uncertainty in countries where EMU data have effectively tracked mCPR in the past. 

Using survey and EMU data available in 2023, we performed out-of-sample validation to assess model predictive performance, comparing performance to the survey-only model as a baseline. It was established that the inclusion of EMU in this manner in addition to the survey model in FPET improved model predictive performance. We saw an improvement across all validation error metrics when predicting both MWRA and UWRA survey observations of mCPR. Additionally, an increase in coverage for UWRA highlights the benefit of using EMU to inform the population for which service statistics have been collected, given in most cases this is for all women. Previously, EMU were solely used to inform estimates for MWRA and subsequently would have had no impact on UWRA estimates when used (\cite{Cahill2021}).

When evaluating the impact of EMU inclusion on 2023 mCPR estimates, we observed maximum impacts of 3.5 percentage points for all women of reproductive age, 4.2 percentage points for married women of reproductive age, and 1.6 percentage points for unmarried women of reproductive age with the use of EMU data. Due to country-level variability, the uncertainty associated with EMUs can be substantial, and in some settings, their inclusion has minimal impact, highlighting one of the model's strengths.

Six country-level case studies were presented in the paper to showcase variation in EMUs across countries, and subsequently, the impact that EMU inclusion has on mCPR estimates and forecasts. There were examples to illustrate the minimal impact EMU will have on model results when there is a recent survey available. Conversely, there were also examples showcasing the use of EMU in situations where there is a survey-absent time period of at least five years. In this case, the use of EMU could have significant impact on mCPR estimates, with impact varying by country and data type. Moreover, each case study provided insight into the impact at a country-level of different levels of observation-specific uncertainty. In some cases, more recent EMUs are associated with higher uncertainty, which reduces their impact on model estimates. In other cases, observation-specific uncertainty is generally low, leading to a greater influence on mCPR estimates. 

The work presented in this paper contributes to empowering countries to track and highlight progress toward their family planning goals in a timely and accurate manner. Updates to the EMU data model and FPET mark a significant advancement in family planning modelling. By extending the EMU data model, we can ultimately help to better inform estimates and forecasts of mCPR for married and unmarried women of reproductive age, aiding countries to more comprehensively monitor progress towards their family planning goals. By providing more accurate and inclusive mCPR estimates, these improvements strengthen the ability of countries to track and achieve their family planning objectives.

\section*{Funding} 
This publication has emanated from research supported by Science Foundation Ireland under Grant number 18/CRT/6049 and by the Bill \& Melinda Gates Foundation (INV-008441). Based on the funding terms of both grants, a Creative Commons Attribution 4.0 Generic License (CC BY 4.0) will be assigned to any Author Accepted Manuscript version arising from this submission.

\section*{Conflicts of interest}
There are no conflicts of interests to declare.

\bibliographystyle{abbrvnat}
\bibliography{references}

\section*{Appendix}

\subsection*{Results for all countries}
  
\textbf{Estimates of mCPR for MWRA, UWRA, and WRA for all countries in the 2023 EMU database.} Solid lines indicate median estimates of mCPR, shaded regions show the 95\% credible intervals. Results in red highlight the use of the survey-only model. Results in blue highlight the use of surveys and EMU. The error bars represent the 95\% confidence interval associated with the data. The purple dashed line marks the year of the most recent survey.

\includepdf[pages=1-,nup=1x3,pagecommand=,width=\columnwidth]{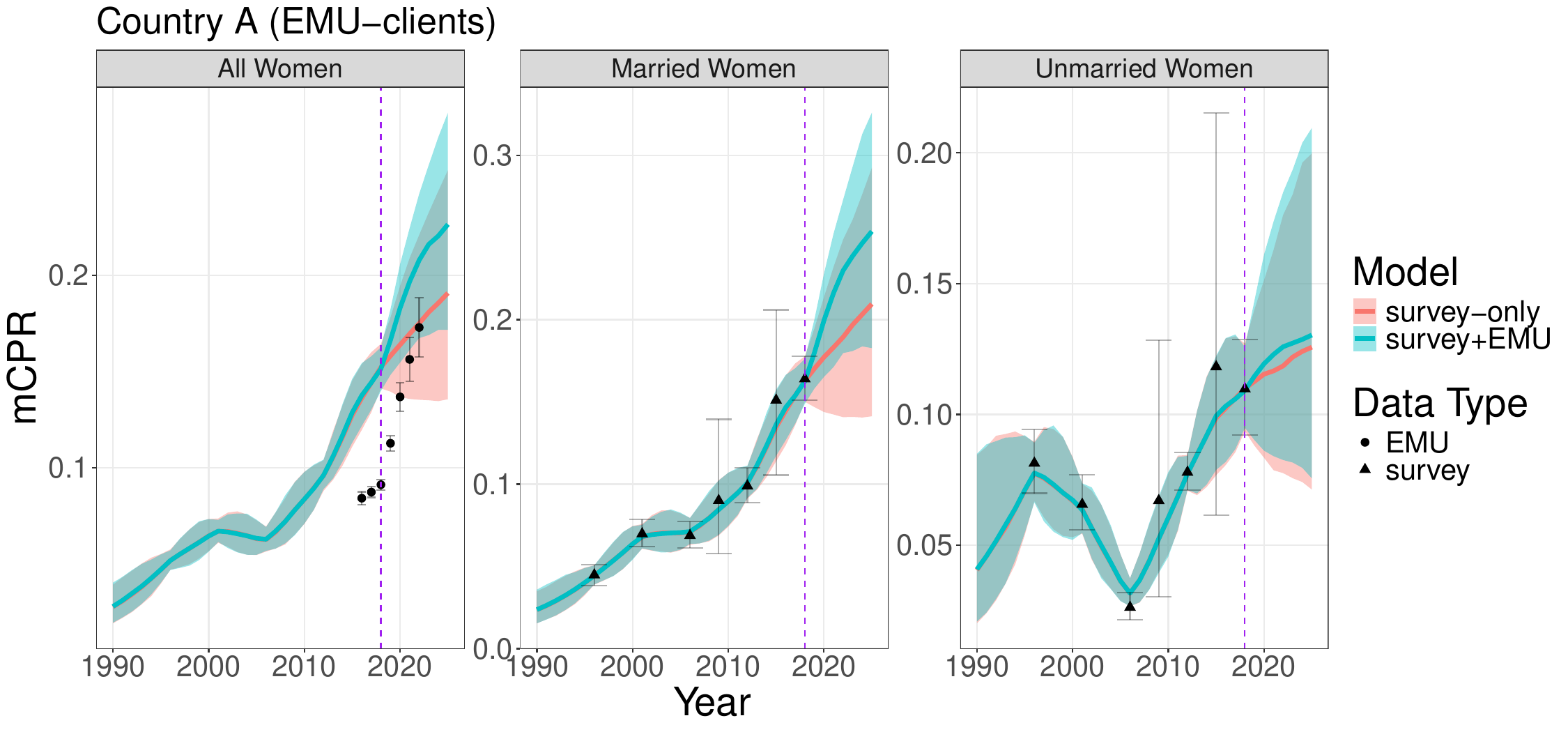}

\end{document}